\documentclass[a4paper,11pt]{article}
\usepackage{pos}

\title{Twisted mass gauge ensembles at physical values of the light, strange and charm quark masses}
\ShortTitle{Twisted mass gauge ensembles at physical values of the quark masses}

\author*[a]{Jacob Finkenrath}
\author[a,b]{Constantia Alexandrou}
\author[a]{Simone Bacchio}
\author[c]{Martha Constantinou}
\author[d]{Petros Dimopoulos}
\author[e]{Roberto Frezzotti}
\author[f]{Karl Jansen}
\author[g]{Bartosz Kostrzewa}
\author[a]{Giannis Koutsou}
\author[e,h]{Giancarlo Rossi}
\author[i]{Carsten Urbach}
\author[j]{Urs Wenger}

\affiliation[a]{Computation-based Science and Technology Research Center, The Cyprus Institute, Nicosia, 20, Constantinou Kavafi Str., Nicosia 2121, Cyprus}
\affiliation[b]{Department of Physics, University of Cyprus, P.O. Box 20537, Nicosia 1678, Cyprus}
\affiliation[c]{Temple University, 1801 N Broad Str., Philadelphia, PA 19122, USA}
\affiliation[d]{Dipartimento  di  Scienze  Matematiche,  Fisiche  e  Informatiche,  Universit\`a  di  Parma  and  INFN, Gruppo  Collegato  di  Parma, Italy}
\affiliation[e]{Dipartimento di Fisica, Universita di Roma "Tor Vergata" and INFN, Sezione di Roma 2, Via della Ricerca Scientifica - 00133 Rome, Italy}
\affiliation[f]{NIC, DESY, Zeuthen, Germany}
\affiliation[g]{High Performance Computing and Analytics Lab, Rheinische
Friedrich-Wilhelms-Universit{\"a}t Bonn, Friedrich-Hirzebruch-
Allee~8, D-53115 Bonn, Germany}
\affiliation[h]{Centro Fermi - Museo Storico della Fisica e Centro Studi e Ricerche Enrico Fermi, Compendio del Viminale, Piazza del Viminiale 1, I-00184, Rome, Italy}
\affiliation[i]{HISKP (Theory), Rheinische Friedrich-Wilhelms-Universit\"at Bonn, Bonn, Germany}
\affiliation[j]{Albert Einstein Center for Fundamental Physics, Institute for Theoretical Physics, University of Bern, Sidlerstrasse 5, CH-3012 Bern, Switzerland}

\emailAdd{j.finkenrath@cyi.ac.cy}

\abstract{Lattice QCD simulations directly at physical masses of dynamical light, strange and charm quarks are highly desirable especially to remove systematic errors due  to chiral extrapolations.  However such simulations are still challenging. We discuss the adaption of efficient algorithms, like multi-grid methods or higher order integrators, within the molecular dynamic steps of the Hybrid Monte Carlo algorithm, that are enabling simulations of a new set of gauge ensembles by the Extended Twisted Mass collaboration (ETMC).   
We  present the status of the on-going ETMC simulation effort that aims to enabling studies of  finite size and discretization effects.  We work within the twisted mass discretization which is free of odd-discretization effects at maximal twist and present our tuning procedure.}

\FullConference{%
 The 38th International Symposium on Lattice Field Theory, LATTICE2021
  26th-30th July, 2021
  Zoom/Gather@Massachusetts Institute of Technology
}

\begin{document}
\maketitle

\section{Introduction}

During the past decade, progress has been made enabling simulations
of twisted mass fermions at physical quark masses at several  lattice spacings less than 0.1~fm  and
volumes as large as 9~fm \cite{ExtendedTwistedMass:2021qui,Alexandrou:2018egz}. These ensembles drive the rich physics program of the Extended Twisted Mass collaboration (ETMC),
which ranges from measurements of quark masses, precision measurements of CKM-matrix elements, hadron spectroscopy and scattering, nucleon structure, semi-leptonic decays  and many other ETMC projects.

In this presentation,  we  overview the progress on the algorithmic and computational
side that made the generation of these ensembles possible. More specifically, we  discuss how to fix the parameters of the simulations, which needs a careful fine tuning procedure 
in order to achieve ${\cal O}(a)$ improvement, how
 to optimize  the multigrid solver DDalphaAMG~\cite{Frommer:2013fsa,Alexandrou:2016izb} and how to improve 
the force computation during the Hybrid Monte Carlo simulations. The use of a three level multigrid 
procedure comes with some limitation for current HPC machines equipped with CPUs that limit scalability.
Furthermore, we present selected results using  the generated statistics and discuss autocorrelations
 using the  physical point ensembles. 

\begin{figure}
 \includegraphics[width=.516\textwidth]{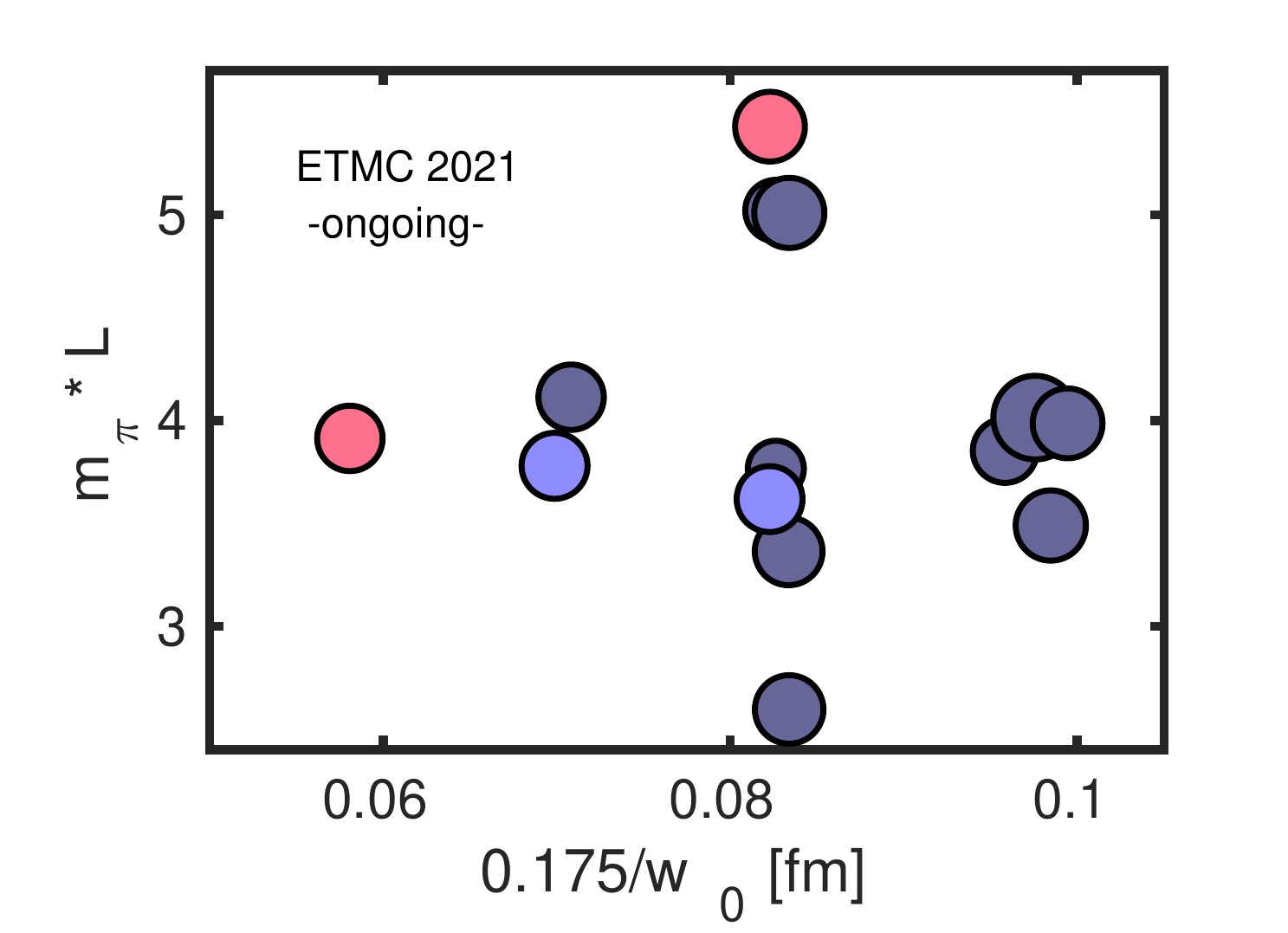}
\includegraphics[width=.484\textwidth]{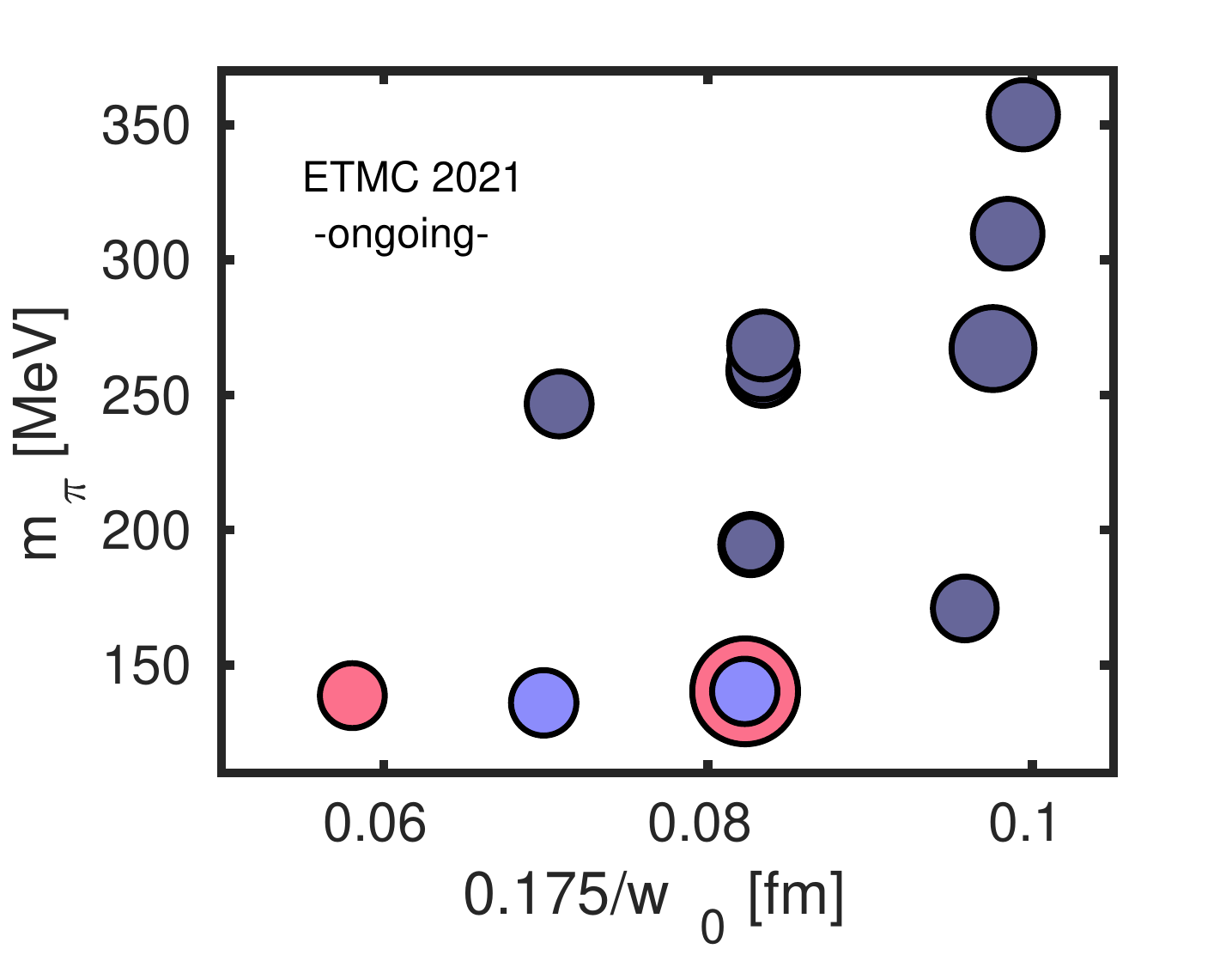}
\caption{Overview of the current  $N_f=2+1+1$  twisted mass clover-improved  ensembles.
The left panel shows the  ensembles according to  their lattice extent in units of  the pion mass, while the right panel shows them according to  the pion mass.
The dark blue circles show the non-physical point ensembles, the light blue already existing physical point ensembles and the red on-going simulations at the physical point.}
\label{fig:ensembleoverview}
\end{figure}

\section{Status of Extended Twisted Mass Collaboration $N_f= 2+1+1$ Simulations}

Within ETMC,   twisted mass clover-improved ensembles has beed generated 
at four different lattice spacings, namely $a\sim 0.093$~fm,  $a\sim 0.08$~fm, $a\sim 0.069$~fm and  $a\sim 0.057$ fm, referred to A-, B-, C- and  D-lattices, respectively. 
We target pion mass ranges between 250 MeV and 135 MeV, with the exception of the A-lattices where ensembles with up to 350 MeV are generated. Ensembles with physical pion masses, also denoted as physical point ensembles, are generated at the B-, C- and D-lattice spacings.
As depicted in Fig.~\ref{fig:ensembleoverview}, all physical point ensembles fulfil $m_\pi \cdot L > 3.6$. A larger volume ensemble is generated at the B-lattice spacing with $m_\pi \cdot L = 5.3$, to be used for finite volume studies. We target statistics exceeding $3000$ MDUs for all ensembles. Parameters are listed in Table~\ref{tab:simparams} and can be also be found in Ref.~\cite{ExtendedTwistedMass:2021qui}.

\begin{table}
        \begin{tabular}{||c|c|c|c|c|c|c|c|c|c||}
                \hline
                  ensemble   & $\beta$ & $c_{\mathrm{SW}}$ &   $\kappa$   &     $V / a^4$     & $a \mu_\ell$ & $a \mu_\sigma$ & $a \mu_\delta$ \\ \hline
                cA211.53.24  & $1.726$ &      $1.74$       & $0.1400645$  & $24^3 \times ~48$ & $~0.00530~$  &   $~0.1408$    &   $~0.1521$       \\
                cA211.40.24  &         &                   &              & $24^3 \times ~48$ & $~0.00400~$  &                &                   \\
                cA211.30.32  &         &                   &              & $32^3 \times ~64$ & $~0.00300~$  &                &                   \\
                cA211.12.48  &         &                   & $0.1400650$  & $48^3 \times ~96$ & $~0.00120~$  &                &         \\ \hline
                cB211.25.24  & $1.778$ &      $1.69$       & $0.1394267$  & $24^3 \times ~48$ & $~0.00250~$  &  $~0.1246864$  &  $~0.131052$ \\
                cB211.25.32  &         &                   &              & $32^3 \times ~64$ & $~0.00250~$  &                &              \\
                cB211.25.48  &         &                   &              & $48^3 \times ~96$ & $~0.00250~$  &                &              \\
                cB211.14.64  &         &                   &              & $64^3 \times 128$ & $~0.00140~$  &                &              \\
                cB211.072.64 &         &                   & $0.1394265$  & $64^3 \times 128$ & $~0.00072~$  &                &    \\
                 cB211.072.96 &         &                   &  & $96^3 \times 192$ & $~0.00072~$  &                &    \\  \hline
                cC211.20.48  & $1.836$ &     $~1.6452$     & $0.13875285$ & $48^3 \times 96$ & $~0.00060~$  &  $~0.106586$   &  $~0.107146$  \\ 
                cC211.125.64  &   &          &     & $64^3 \times 128$ & $~0.00125~$  &   &    \\ 
                cC211.06.80  &   &          &     & $80^3 \times 160$ & $~0.00060~$  &   &    \\ \hline
                cD211.17.64  & $1.900$ &     $~1.6112$     & $ 0.137972174$ & $64^3 \times 128$ & $~0.00170~$  &  $~0.087911$   &  $~0.086224$   \\ 
                cD211.054.96  &               &     &               & $96^3 \times 192$ & $~0.00054~$  &   &   \\ \hline
        \end{tabular}
    \caption{Simulation parameters of the $N_f=2+1+1$ twisted mass clover-imporved ensembles.}
    \label{tab:simparams}
\end{table}

\subsection{Twisted mass fermion action}

The $N_f=2+1+1$ ensembles are generated using the  Iwasaki gauge action for the pure gauge, the  $N_f=2$ mass-degenerate twisted mass fermion action with a clover term
for the light quarks and the  $N_f= 1+1$ non-degenerate twisted mass fermion action for the strange and charm quarks. The non-degenerated twisted mass operator is given in the 	heavy doublet flavor space by
\begin{equation}
   D(\kappa,c_{sw},\mu,\epsilon)  =  D_{W}(\kappa,c_{sw}) \otimes {1} + i \mu_\sigma \gamma_5 \otimes \tau_3  - \mu_\delta \otimes \tau_1 = \begin{bmatrix}
                                                                            D_W + i \gamma_5 \mu_\sigma & -\mu_\delta \\
                                                                            -\mu_\delta  & D_W - i \gamma_5 \mu_\sigma
                                                                           \end{bmatrix} 
                                                                       \label{eq:NDtm}
\end{equation}
with $D_W$ the clover improved Wilson Dirac operator, $\kappa$ the Wilson hopping parameter, $c_{SW}$ the clover parameter and $\mu_\sigma$ and $\mu_\delta$ the 1+1 twisted mass parameters~\cite{Frezzotti:2003xj}. Note that taking $\mu_\delta  = 0$  the mass-degenerate twisted mass operator
in flavor space is recovered.
Twisted mass fermions have several advantages. Lattice artefacts of odd power in $a$ can be removed
if the Partially Conserved Axial Current (PCAC) mass is tuned to zero, i.e.~$m_{PCAC}(\kappa) \longrightarrow  0$   \cite{Frezzotti:2003ni}. This can be done
by fine tuning the bare Wilson quark mass parameter  $\overline{m}=0.5/\kappa-4$ to its critical mass $\overline{m}_{crit}=0.5/\kappa_{crit}-4$.
Moreover, in the case of a finite twisted mass value, the operator $D$  is non-singular. For the squared operator one gets in fact
\begin{equation}
  D^\dagger D = D_{W}^\dagger D_{W} + \mu^2 \, ,
\end{equation}
with $\mu$ playing the role of an infra-red cut-off for the eigenvalues of $D$.
While this guarantees convergence of iterative methods, such as the conjugate gradient solver,
the twisted mass term breaks isospin symmetry. This results in a mass-splitting within the pion triplet. The neutral pion mass is shifted with respect to the charged pion mass, which is given in next to leading order (NLO)  chiral perturbation theory~\cite{Sharpe:2004ny} by 
  \begin{equation}
a^2 (m_{\pi^0}^2 - m_{\pi^\pm}^2) = -c_0 \cdot a^2   \,.
 \label{eq:pionmassgap}   
  \end{equation}
This cut-off effect, if large, can give neutral pion mass of zero at finite quark masses and trigger a phase transition. For values of the  neutral pion mass close to zero, tuning the PCAC mass to zero becomes notoriously difficult and such Monte Carlo simulations of twisted mass fermions become impossible.
This prevented simulation at physical pion masses without a clover term.

Including a clover term,  reduces lattice artefacts which decreases the mass gap between the charged and neutral pion given in  Eq.~\eqref{eq:pionmassgap} to such an extent as to make simulations at values of $a\sim 0.09$ possible~\cite{ETM:2015ned}.
For our setup we use  1-loop tadpole boosted perturbation theory \cite{Aoki:1998ph} to fix the value of the  $c_{SW}$ parameter, given by
\begin{equation}
c_{SW} = 1 + 0.113(3) \frac{6}{\beta \langle P \rangle}
\label{eq:csw}
\end{equation} 
with $\langle P \rangle$ the plaquette.

\subsection{Parameter tuning}

In order to simulate $N_f=2+1+1$ twisted mass fermion ensembles at the physical point, 
we need to set the bare parameters for our action. The complete set of parameters is 
\begin{equation}
 \{ \beta , c_{SW}, \kappa, \mu_\ell, \mu_\sigma, \mu_\delta \} \; .
\end{equation}
Using an estimated initial guess, we can pre-select the value of the gauge coupling $\beta$ 
that fixes $c_{SW}$ via Eq.~\eqref{eq:csw}. 
The bare-mass parameters of the twisted mass action
that are left, namely
\begin{equation}
\{ \kappa, \mu_\ell, \mu_\sigma, \mu_\delta \}
\end{equation}
require  careful fine tuning  in order
to guarantee $\mathcal{O}(a)$ improvement.
This can be done by tuning the bare Wilson quark mass $\kappa$ towards its critical value by requiring 
\begin{equation}
\frac{Z_A m_{PCAC} ( \kappa, \mu_{\ell}, \mu_\sigma , \mu_\delta) }{\mu_\ell} < 0.1,
\label{eq:PCAC_condition}
\end{equation}
with $Z_A$ the axial renormalization factor.
Note that the PCAC mass depends also on the heavy quark parameters $ \mu_\sigma$ and $\mu_\delta$  of the non-degenerated  twisted mass operator. Due to this dependence on the heavy quark pair, charm and strange, we utilise Osterwalder Seiler (OS) fermions \cite{Frezzotti:2004wz} to set the strange and charm quark mass. We then match the non-unitary setup of the OS-fermions with the unitary setup with non-degenerate twisted mass fermions.
This results in three tuning conditions for the heavy quark parameters.
The first two are given by
\begin{equation}
 C_1 = \frac{ \mu_{c}^{OS} } {\mu_{s}^{OS} } = 11.8 \qquad \textrm{and} \qquad C_2 = \frac{ m_{D_s}^{OS} } {f_{D_s}^{OS} } = 7.9
\end{equation}
that set the strange and charm quark mass parameters in the OS sector.
The full matching between OS and non-degenerate twisted fermions is finally done using  the kaon mass $m_K$ by requiring
\begin{equation}
 m_K^{OS}(\mu_{s}^{OS} ) \equiv m_K^{ND}(\mu_\sigma,  \mu_\delta), 
\end{equation}
which basically utilizes the renormalization factor ratio $Z_P/Z_S$ and yields to
the determination of $\mu_\sigma$ and $\mu_\delta$ via the relation
 $\mu_{c,s}^{OS} = \frac{1}{Z_P} \left(\mu_\sigma \pm \frac{Z_P}{Z_S} \mu_{\delta}\right)$ .

\section{Parameter tuning}

\begin{figure}
\includegraphics[width=0.33\textwidth]{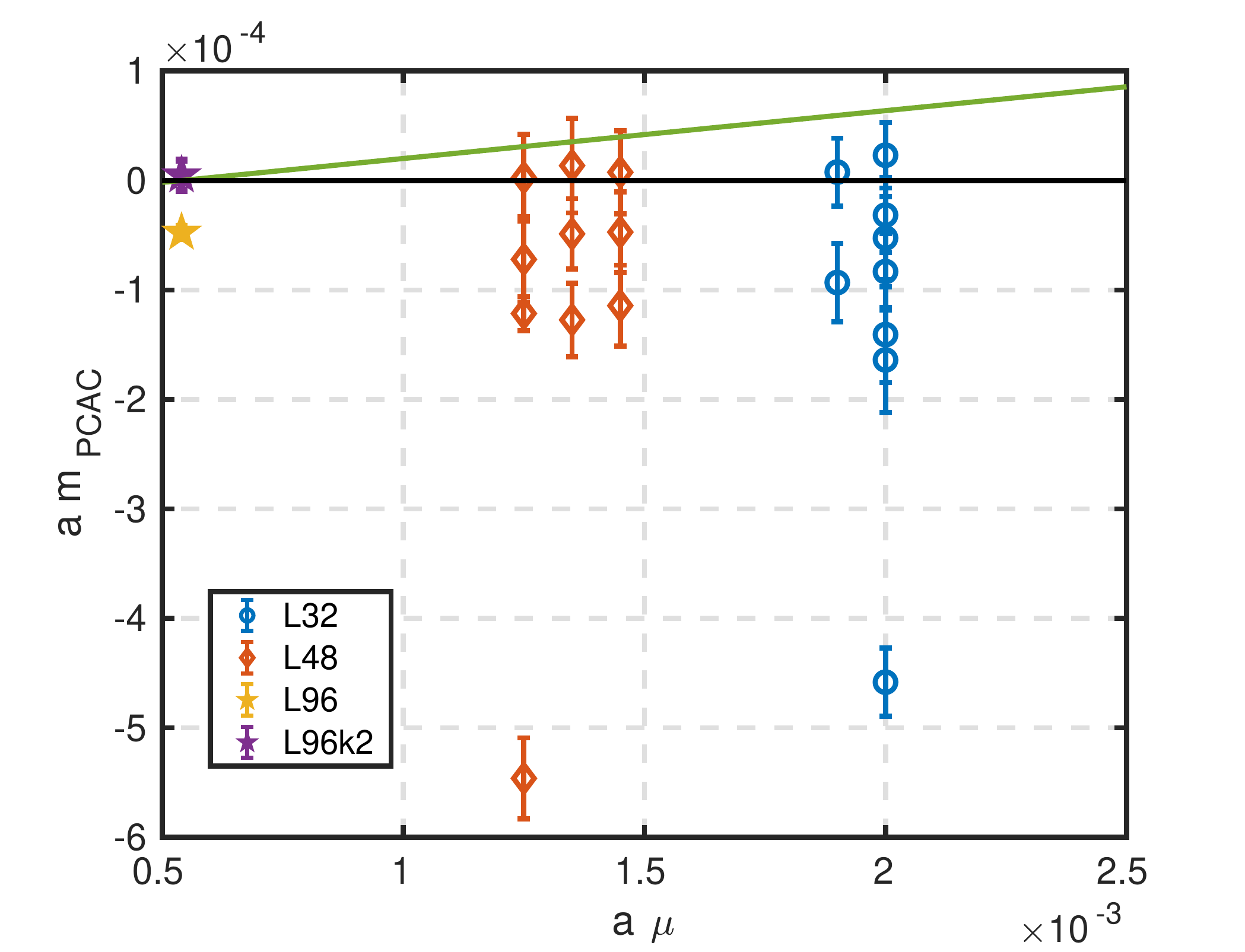}
\includegraphics[width=0.33\textwidth]{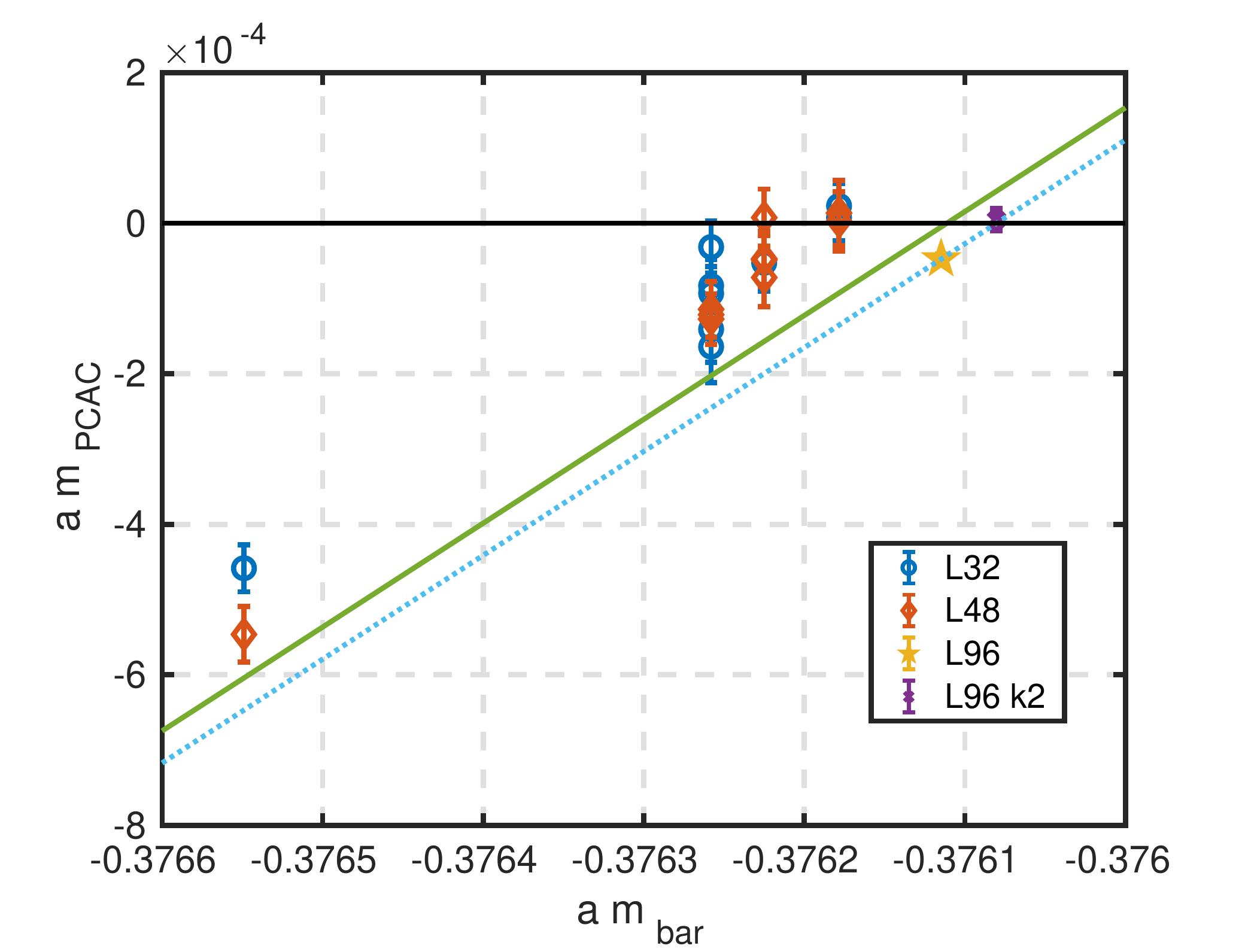}
\includegraphics[width=0.33\textwidth]{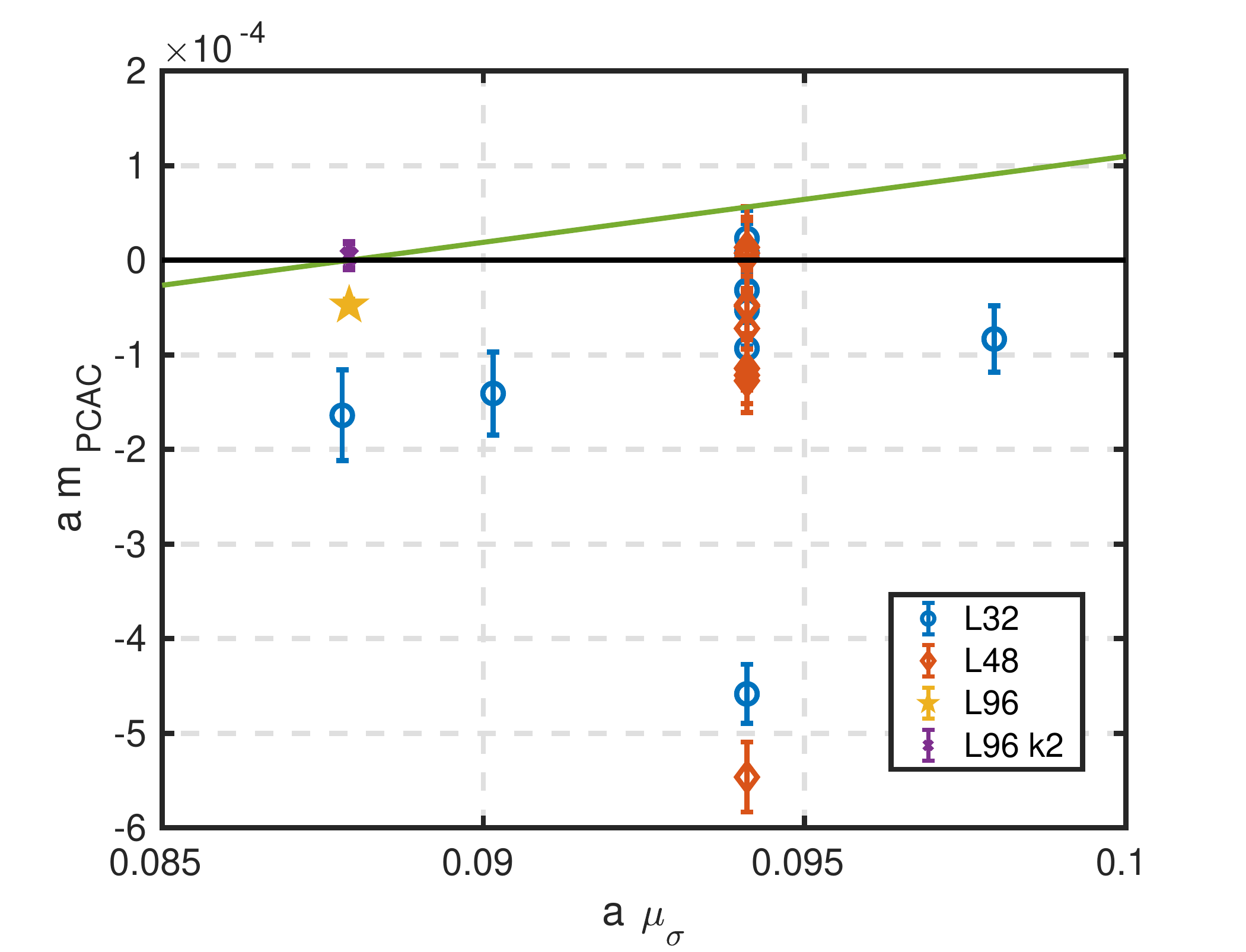}
\caption{The tuning procedure employed for the D-lattice spacings using reweighting on two tuning ensembles. We depict the dependence of the PCAC mass on the light twisted mass parameter  $a \mu_\ell$ (left), on the Wilson bare quark mass $a\bar{m}$ (middle)  the heavy quark parameter $\mu_{\sigma}$ (right).}
\label{eq:tuning_Densem}
\end{figure}

For the  tuning of the twisted mass parameters, we iterate the following steps: i)  tune $\kappa$ at constant $\{\mu_\ell, \mu_\sigma, \mu_\delta \}$; ii) change $\mu_\ell$ and retune $\kappa$; iii) tune $\{\mu_\sigma, \mu_\sigma\}$ and retune $\kappa$; iv) reiterate until stability is reached.
This procedure requires for the tuning of the A-, B- and C- ensembles  to generate for each steps roughly two Markov chains with around 500 MDU,
see \cite{Alexandrou:2018egz} for a detailed discussion for  the case of the  B-lattice spacing.
For the  D-lattice we modify the tuning procedure by making use of reweighting for all parameters  $\{ \kappa, \mu_{\ell}, \mu_\sigma, \mu_\delta \}$ \cite{Hasenfratz:2008fg,Finkenrath:2013soa,Alexandrou:2020bkd}.
This reduces the number of ensembles needed for the tuning, namely to one per iteration circle.

With the help of reweighting by changing the bare parameters by 1\% to 5\% we could give an 
estimate of the first derivatives of  $a m_{PCAC} ( \kappa, \mu_{\ell}, \mu_\sigma, \mu_\delta)$. This quantity 
was estimated in the case of the tuning procedure of the A-, B- and C-lattice spacings via generation of several ensembles.

For the tuning of the D-lattice spacing we generate two different ensembles at  twisted mass values $a \mu = 0.002$ and $a \mu = 0.00125$
and  volumes of $V = 32^3\times 64$ and $V = 48^3\times 96$, respectively.  We employ the fit Ansatz
\begin{equation}
f( \overline{m},\mu_\ell, \mu_\ell )= c_0 + c_1 \overline{m} + c_2 \mu_\ell + c_3\mu_\ell
\end{equation}
with  bare quark mass $\overline{m}=1/2/\kappa -4$. We find for the fit coefficients
$c_0 = 0.51(2)$, $c_1=1.37(6)$, $c_2= 0.04(2)$ and $c_3=0.012(4)$.
The resulting uncorrelated fit with $\chi^2 = 11.2$ is shown in Fig.~\ref{eq:tuning_Densem}, where we  
include all available data points generated by reweighting resulting in 19 degrees of freedom.
The critical mass parameter at physical light quark mass of $a \mu = 0.00054$ 
is given by $\kappa_{crit}^{D,(0)} = 0.137973465$.
Based on this estimate we generate 500 MDUs at our the physical point with $a \mu=0.00054$, leading to a slightly negative PCAC mass
of $-4.4(7)10^{-5}$. This does meet our criterion and thus we proceed to retune $\kappa$ using our estimate for the slope $\partial f/\partial \overline{m} = 1.37 $
such that our final estimate for the critical Wilson mass parameter is given by $\kappa_{crit}|_{\beta=1.9} =0.137972174 $ .

\begin{figure}
\includegraphics[width=1.0\textwidth]{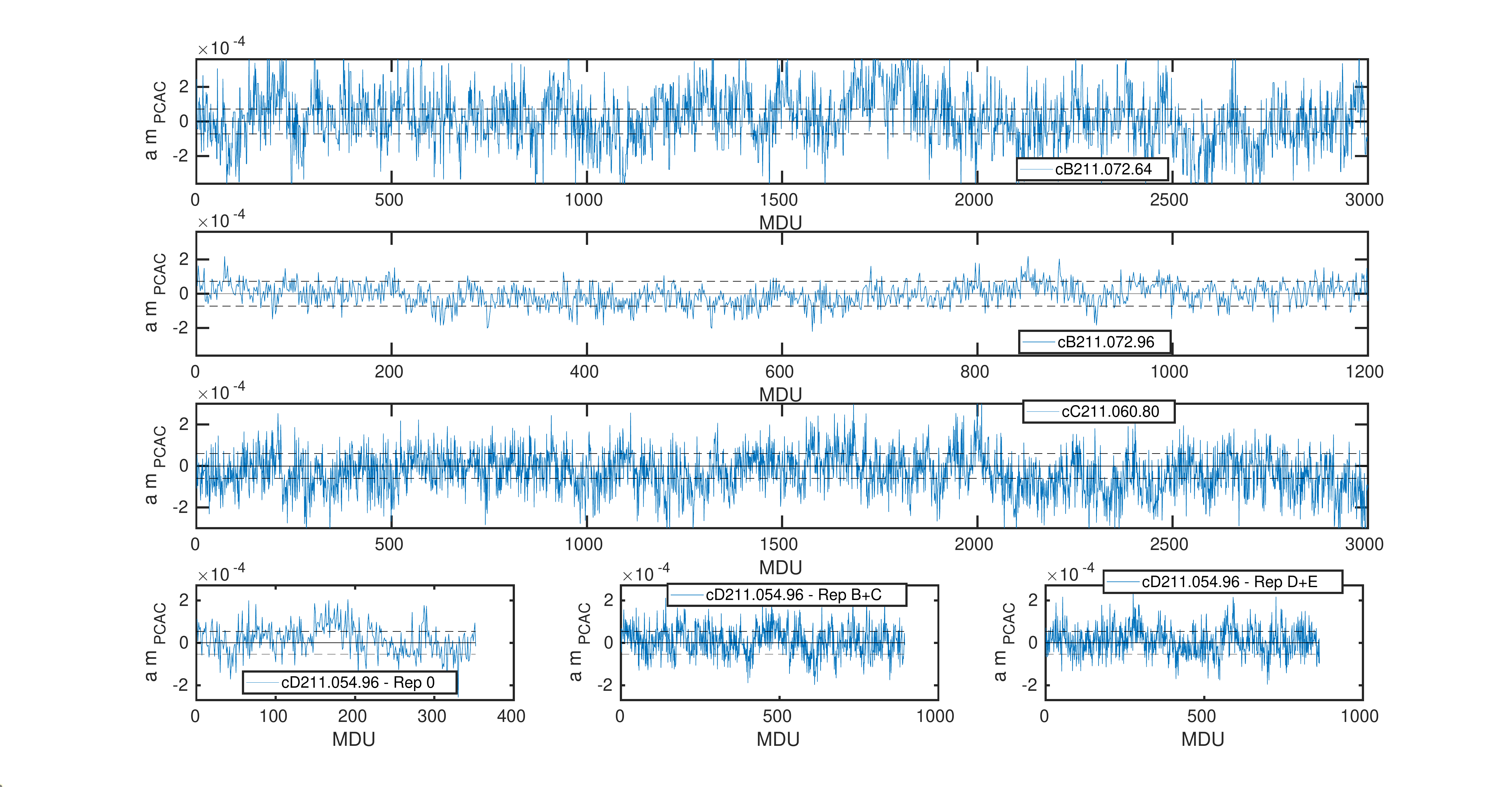}
\caption{The history of the PCAC mass for the physical point ensembles is plotted in units of molecular dynamics (MDUs) showing from top to bottom  the  cB211.072.64, cB211.072.96, cC211.060.80 and the two replicas of cD211.060.80 ensembles.  }
\label{fig:PCAC}
\end{figure}

With this tuning procedure, that it is also outlined in ref.~ \cite{Alexandrou:2018egz} for the B-lattice spacing, we were able to successfully tune towards critical Wilson mass, achieving $\mathcal{O}(a)$-improvement. Indeed as seen in Fig.~\ref{fig:PCAC}, all ensembles fulfil the condition of eq.~\ref{eq:PCAC_condition} with $m_{PCAC}/\mu < 4\%$ or better.

\section{Simulation setup}

For the generation of the ensembles listed in Table~\ref{tab:simparams}, we use our open source software suite \textit{tmLQCD} \cite{Jansen:2009xp,Abdel-Rehim:2013wba,Deuzeman:2013xaa},
which implements an optimised Hybrid Monte Carlo algorithm enabling the use of twisted mass operators of  Eq.~\eqref{eq:NDtm}
and improved gauge action with rectangular loops.
For the molecular dynamics, we are using,  for the mass-degenerated light quark doublet, even-odd Hasenbusch mass preconditioning with masses $\{\rho_0=\mu_\ell,\rho_1,\ldots,\rho_N\}$, while for the heavy quark doublet, even-odd rational approximation of the square root of the non-degenerate twisted mass operator $\hat{Q}^2_{ND}=D_{ND,eo} D_{ND,eo}^\dagger$ is used \cite{Urbach:2005ji}.
Thus, the Boltzmann weight of our setup is given by
\begin{eqnarray}
 W(U) &=& Z^{-1}\textrm{exp}\Bigg{\{} - \beta S_{iwa}(U) + \textrm{Tr}\, \textrm{ln} \{ W_{oo}(\rho_0,\mu_\sigma) \} -  \sum_{j=1}^N \phi^\dagger \left[\frac{q_i}{\hat{Q}^2_{ND} + \mu_i} \right] \phi \notag\\
  &&  - \prod_{j=1}^{N-1} \eta_j^\dagger (1 + \Delta^2 \rho_{j-1,j} (\hat{Q}^2+ \rho_{j-1}^2 )^{-1} \eta_j - \eta_N^\dagger ( \hat{Q}^2 + \rho^2_N )^{-1}\eta_N \Bigg{\}}\notag .
\end{eqnarray}
For a more detailed discussion see Appendix A of Ref.~\cite{ExtendedTwistedMass:2021qui}. 
Our simulation code $\textit{tmLQCD}$ provides currently a link to the algebraic multigrid solver library DDalphaAMG, which provides routines for the twisted mass operator \cite{Alexandrou:2016izb} and for the non-degenerated twisted mass operator \cite{Alexandrou:2018wiv} that can speed-up the smallest shifts within the rational approximation. Additionally, we have employed the mixed-precision linear solver provided by the software package \textit{QPhiX} \cite{Joo:2013lwm,Joo:2016lwm,Joo:2015lwm,Heybrock:2014iga,QPhiX-github}, which can be utilized for the larger mass shifts, where the algebraic multigrid solver becomes less effective.

\subsection{Application of HMC with multigrid solver}

\begin{figure}
\includegraphics[width=0.336\textwidth]{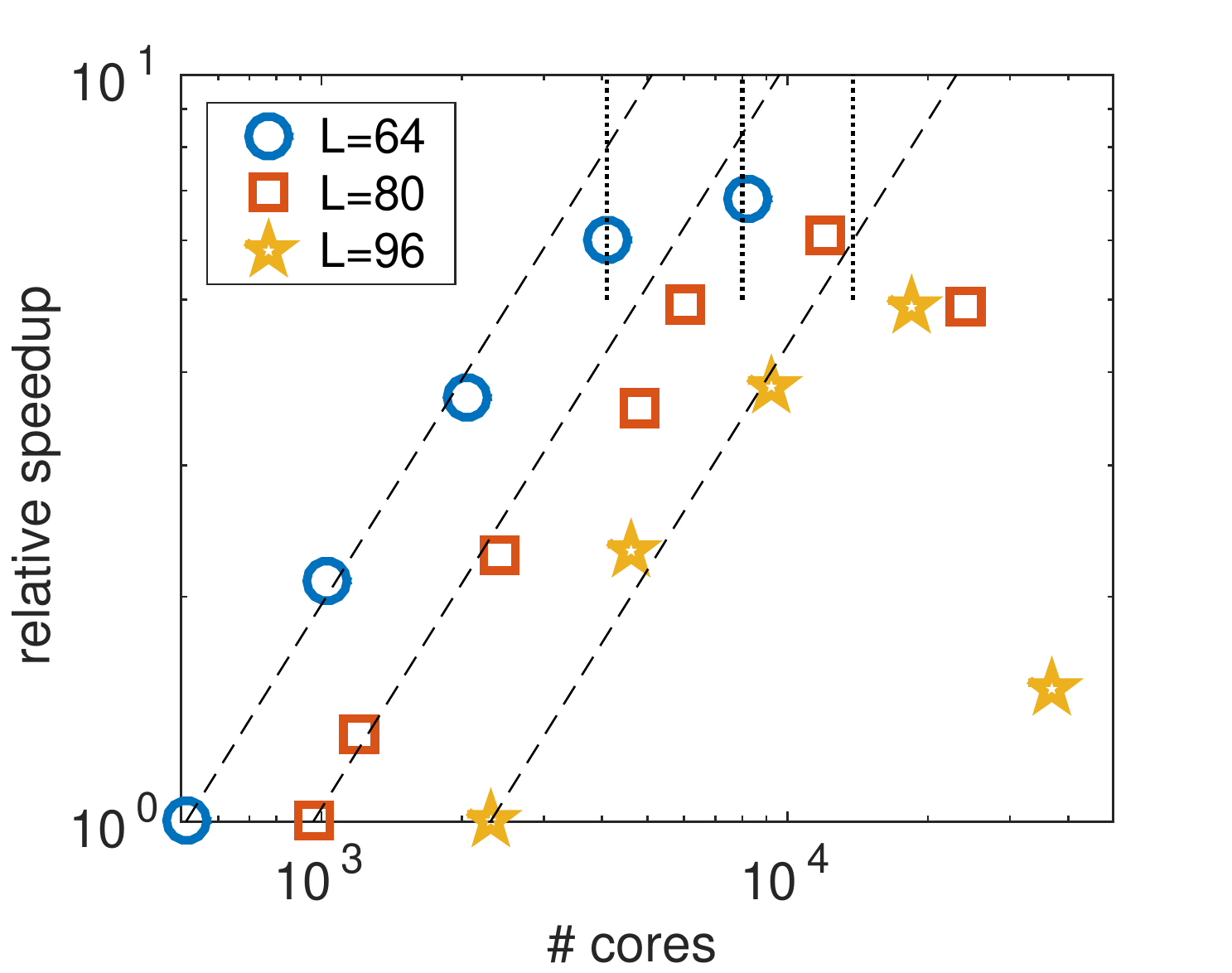}
\includegraphics[width=0.296\textwidth]{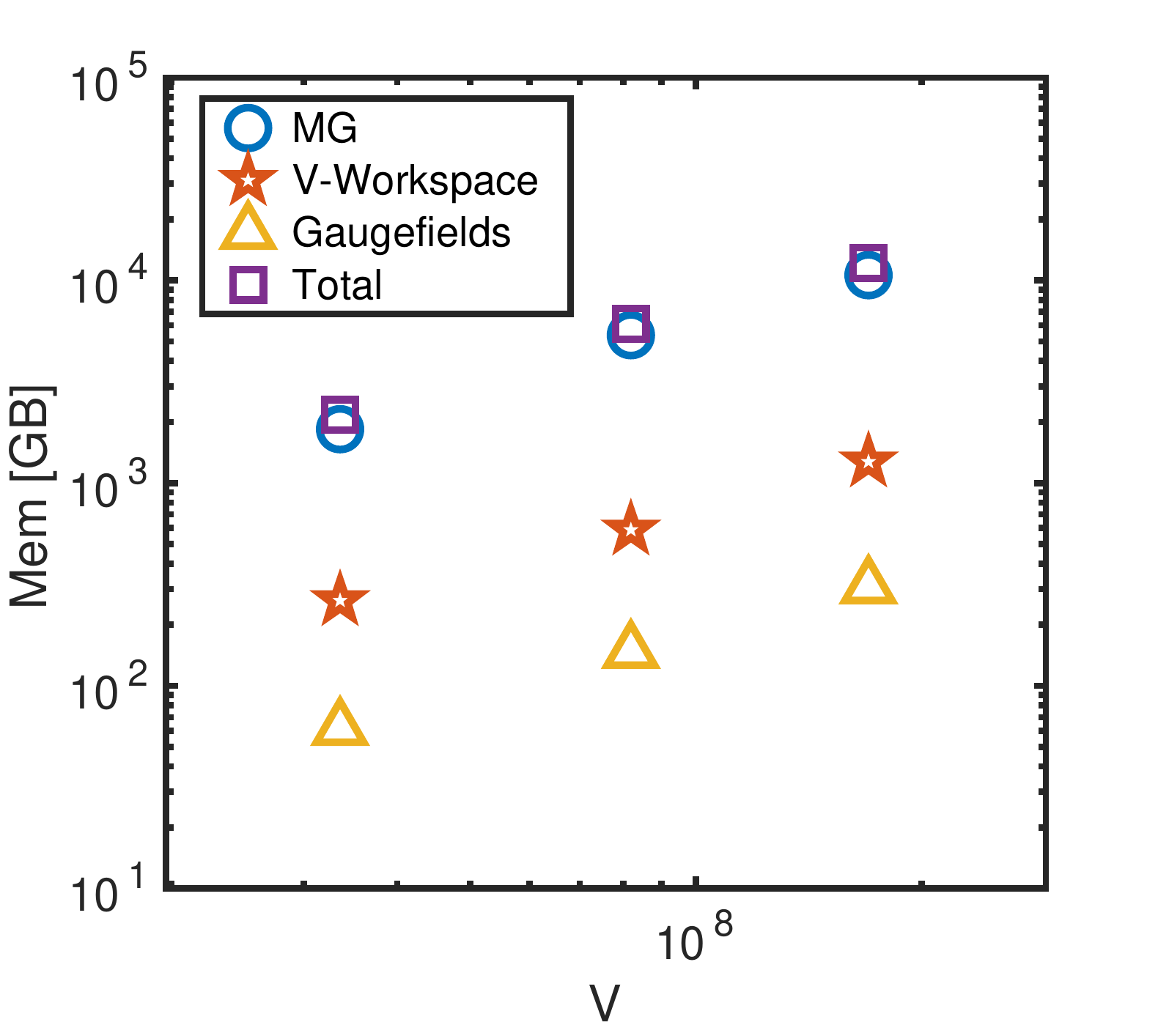}
\includegraphics[width=0.347\textwidth]{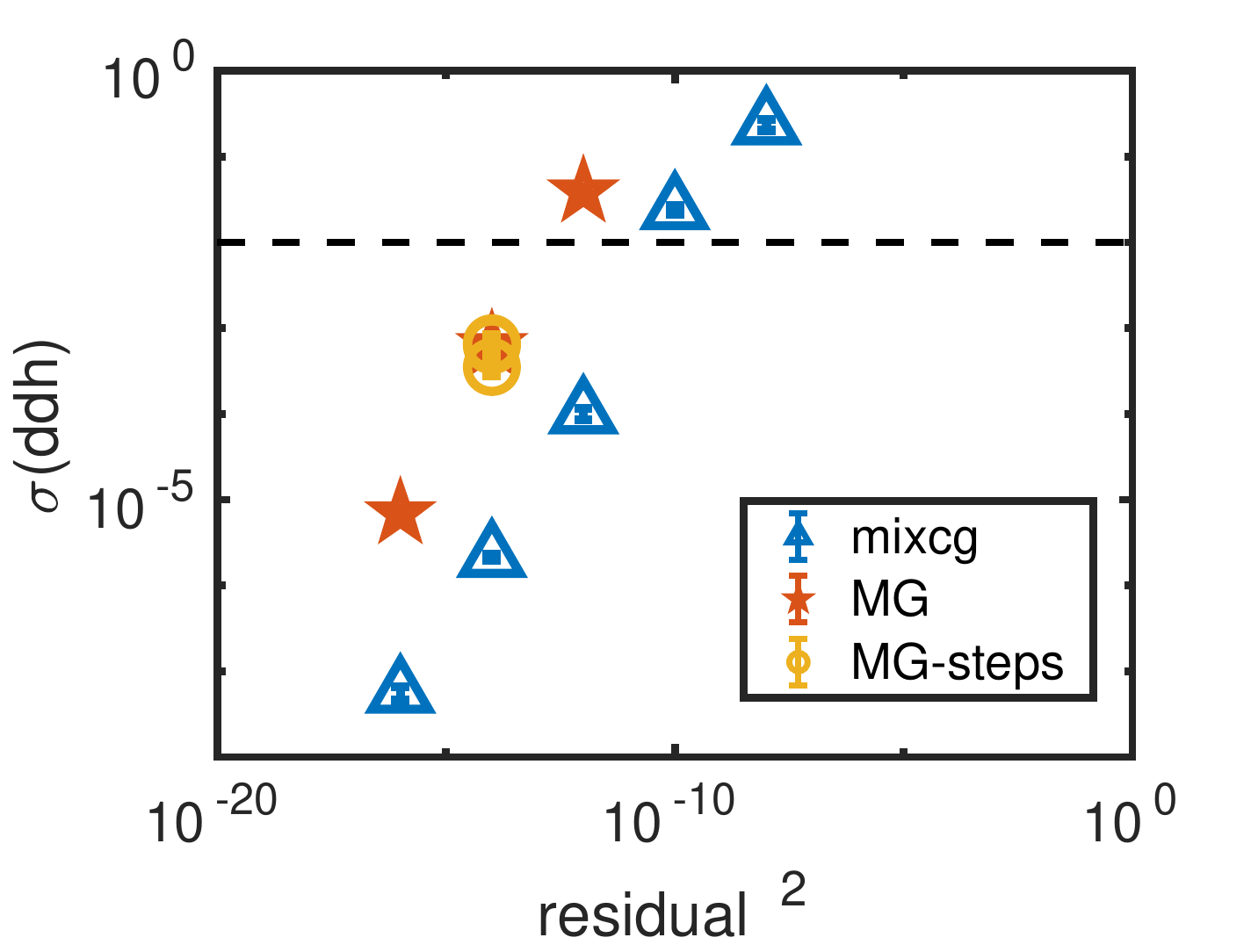}
\caption{The  requirements of a three level multigrid method within HMC simulations are shown.The  left panel shows the scalability for three different volumes on SuperMUC-NG. The middle panel shows the memory requirements and the right panel illustrates the reversibility violation as a function  of the solver residual.}
\label{fig:MGlimits}
\end{figure}

For all simulations of the physical point ensembles,  we take advantage of the highly reduced computational costs
of using multigrid solvers within the force calculation, see e.g.~\cite{Alexandrou:2016izb}.
The use of multigrid solver within the integration of Hamiltons molecular dynamics requires some additional care and comes with some limitations. Namely, the scalability on HPC systems, such as SuperMUC-NG, is limited by the volume of the coarsest grid within the multigrid procedure.
This limits the strong scaling window, as depicted in Fig.~\ref{fig:MGlimits}, breaking down for a three-level MG method for a lattice of size $V=64^3\times 128$ at around 80 Skylake nodes, and for a lattice of size $V=96^3\times 192$ at around 420 Skylake nodes. This results in a roughly scaling of the upper bound of the strong scaling window by $L^3$ or $V^{3/4}$ and limits the maximal effective parallelisation of our HMC.

The lower bound of the working window using a multigrid solver is determined by memory requirements.
In fact prolongation and restriction operators required for projecting iteration vectors from level to level and building up the coarser operators,
need allocation of $\mathcal{O}(20)$ full vectors,  a number which scales with the size of the volume. On SuperMUC-NG with 192 GB RAM per node this introduces a hard limit for the minimal parallelisation, given by about 16 nodes for lattice size of $V=64^3\times128$ and increasing to 80 nodes for a volume of $V=96^3 \times 192$. 

The usage of a multigrid solver not only limits the scalability window, but can also compromise the correctness of the HMC sampling. By reusing and updating the coarse grid operators and the corresponding prolongation and restriction operators from previous integration steps, the reversibility criterium,
which is needed to fulfil detailed balance, is violated. In order to assess the magnitude of the effect, we studied reversibility violation within the HMC using a test volume with size $V=32^3\times 64$ and measuring the variance of $\delta \Delta H$ for different solvers. $\delta \Delta H$ is given by the difference of the Hamiltonian at the beginning of the trajectory with the Hamiltonian integrated to $\tau=1$ and integrated back to the start. As discussed in Ref.~\cite{Urbach:2017ate} for variances below 0.01 no deviations from the expectation values, such as the plaquette, are found. To match the same precision in $\sigma^2(\delta \Delta H)$ as the mixed-precision solver of QPhiX, the square residual of the multigrid solver needs to be two orders of magnitudes smaller. For all our physical point ensembles, we checked reversibility and find that for the selected solver criteria (see \cite{ExtendedTwistedMass:2021qui}), $\sigma^2(\delta \Delta H)$ is below the bound of set in Ref.~ \cite{Urbach:2017ate}.

\subsection{Nested force gradient integrator}

\begin{figure}
\begin{center}
\includegraphics[width=0.33\textwidth]{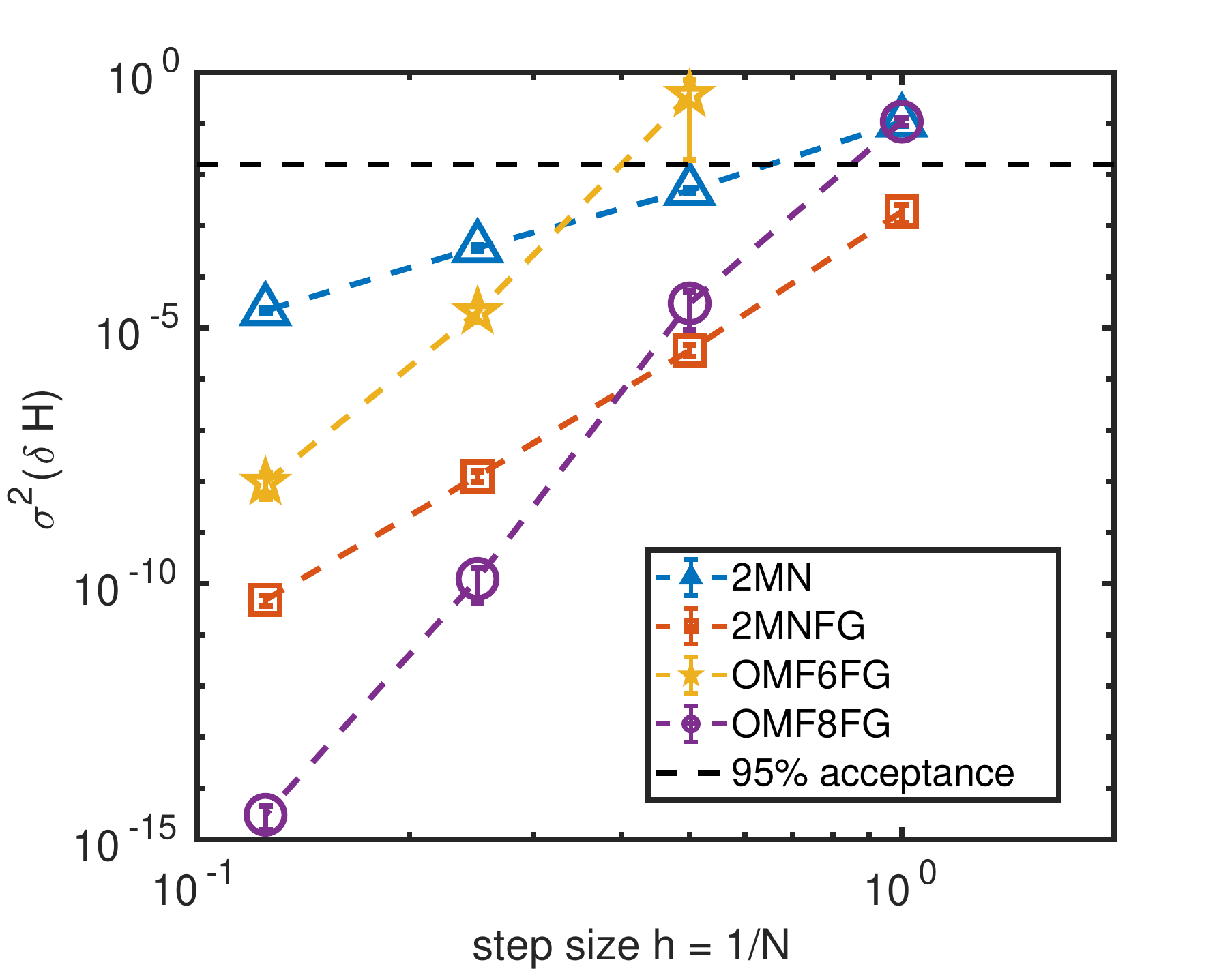}
\end{center}
\caption{Comparison of different numerical integrators. The variance 
of the energy violation is plotted as a function of   the step size for a test ensemble of volume $V=24^3\times 48$.}
\label{fig:integratorstudy}
\end{figure}

The computational cost of molecular dynamics scales with the number of integration steps.
At constant acceptance rate it follows 
  $\textrm{cost} \propto V^{1+\frac{1}{2 n}}$
with $n$ the order of the selected integrator. Thus higher order schemes have a better volume scaling.
A class of various different simplectic reversible integration schemes are discussed in Ref.~\cite{Omelyan:2003}.
 This includes schemes improved with force gradient terms. For example, the second minimal norm scheme can be extended to fourth order 
 \begin{equation}
  \Delta(h) = e^{h \frac{1}{6} \hat{B}} e^{h \frac{1}{2} \hat{A}} e^{h \frac{2}{3} \hat{B}-\frac{1}{72}h^3 \hat{C}}  e^{h \frac{1}{2} \hat{A}} e^{h \frac{1}{6} \hat{B}}
 \label{eq:2mnfg}
 \end{equation}
by including the force gradient term $C$, which is given by
 $C = 2 \sum^{V,3}_{x=1\nu=0} \frac{\partial S}{\partial U_\nu (x)}\frac{\partial^2 S}{\partial U_\nu (x) \partial U_\mu (x)}$ .
The additional second derivate term, needed in the force gradient term, can be approximated by an additional force term,
as outline in Ref.~\cite{Yin:2011np}. This not only reduces the cost of the calculation of one force calculation but
also simplifies the application by implicitly taking care of cross-terms between different parts of the actions.
It turns out that the force gradient improves the minimal norm scheme of Eq.~\ref{eq:2mnfg} and  outperforms for larger volumes the 
other integrators, as shown in  Fig.~\ref{fig:integratorstudy} for a $L=24$ lattice. Note that to tune the nested integrator
setup we minimise the cost function at constant acceptance rate using an effective model for the higher order
terms in line with Ref.~\cite{Clark:2010qw}.

\subsection{Computation costs}

\begin{figure}
\begin{center}
\includegraphics[width=0.6\textwidth]{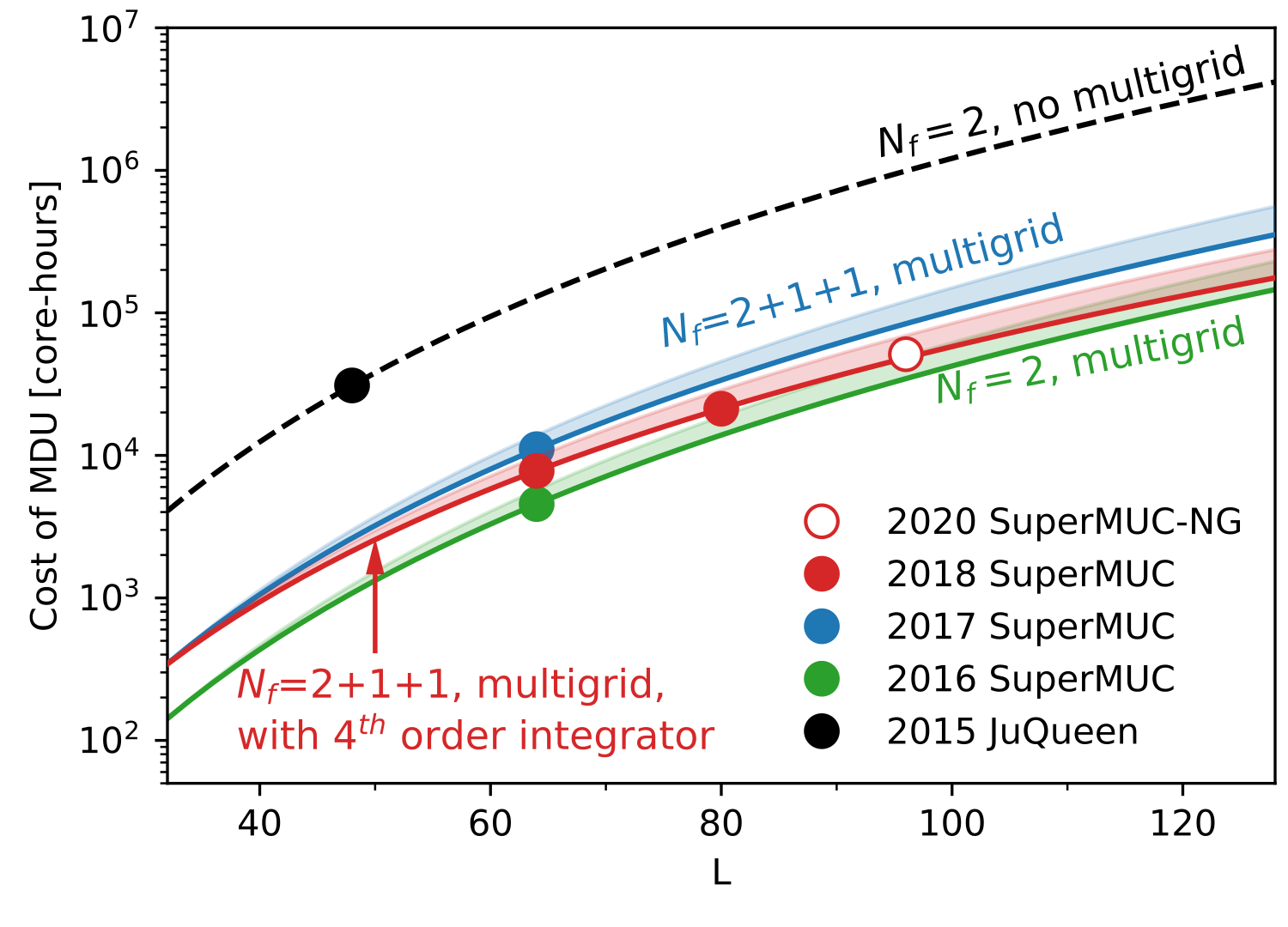} 
\end{center}
\caption{The  computational cost of a trajectory for $N_f=2+1+1$ simulations as a function of the lattice spatial extent $L$.  }
\label{fig:comp}
\end{figure}

A summary of the improvements and the achieved reduction in computational costs per HMC trajectory
 are shown in Fig.~\ref{fig:comp}.
By  enabling DDalphaAMG within the HMC, there is a reduction of  the computational cost per trajectory
by more than one order of  magnitude~\cite{Alexandrou:2016izb}.
In addition, the adaptation of DDalphaAMG to the non-degenerate twisted mass operator
speeds up the non-degenerated sector, as discussed in Ref.~\cite{Alexandrou:2018wiv}.
The use of higher order integrators has given further improvements, especially going to larger volumes, such as $L=96$~\cite{Shcherbakov:2015hhg}.
Note that there is still space for further improvements, for example, by coarse level improvements, which are currently under investigation. Moreover, a multi right handside version
of DDalphaAMG is available, which, however to utilize, would require larger refraction of the force computation~\cite{Shuhei:lattice}. 
We are investigating how this potentially could be adapted within a lattice QCD python API \cite{Bacchio:lattice}, which is currently under development.

\subsection{ Autocorrelations at the physical point }

One major unsolved challenge in lattice QCD with periodic boundary conditions is adequate sampling of different topological sectors 
at very fine lattice spacings \cite{Schaefer:2010hu}. We have monitored the gauge definition of the topological charge $Q$ at gradient flow time $t_0$.
As expected for our range of lattice spacings between $a\sim 0.057$ fm to $a\sim 0.082$ fm, the topological charge is fluctuating well between topological sectors. Moving towards finer lattice spacing, we have currently indication for an increase of the autocorrelation time, hinting at the fact that
for simulations below $a<0.045$ fm further algorithmic improvements will be needed in order to sample the topological charge properly within 3000 MDUs.

\begin{figure}
\includegraphics[width=0.99\textwidth]{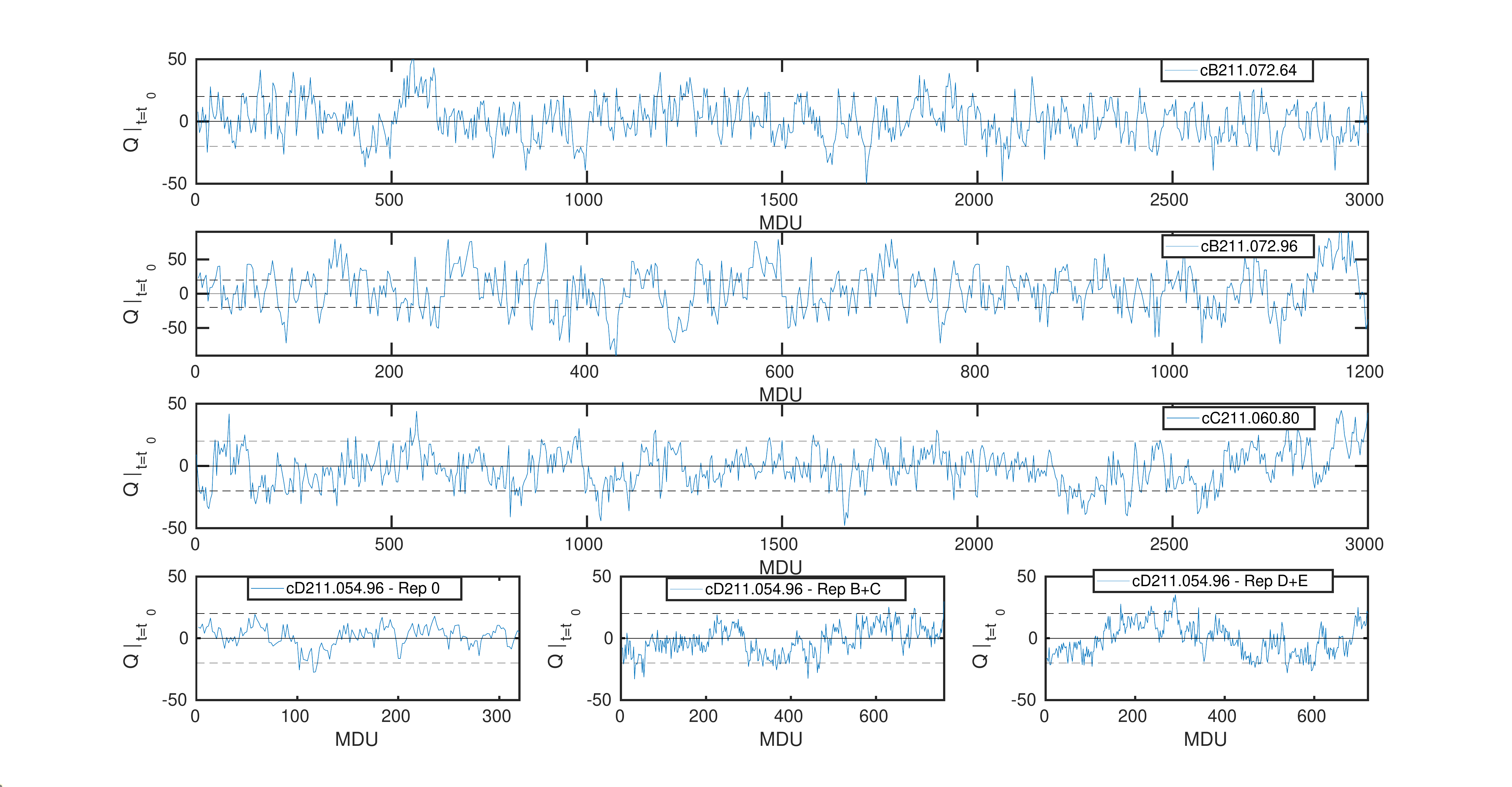}
\caption{The history of the topological charge $Q$ at gradient flow time $t_0$ of the physical point runs plotted versus  the molecular dynamics (MDUs). We show from top to bottom  cB211.072.64, cB211.072.96, cC211.060.80 and two different replicas for cD211.060.80. }
\end{figure}

\section{Conclusions}

With the current ensembles we are able to study finite volume and lattice spacing  artefacts
directly at the physical point. Using the three ensembles cB211.072.64, cC211.060.80 and
cD211.054.96, we can take the continuum limit, and with the two ensembles at the B-lattice spacing
we can study finite volume effects.
In the future, we are planing to simulate at  larger volumes and explore new approaches for enabling simulations at lattice spacings smaller than 0.05~fm, which are currently
limited by the critical slowing down of the algorithms and their scalability with the volume. Due to the behavior of the strong scaling window
of our multigrid solver DDalphaAMG, the real time per trajectories will further increase resulting in longer
generation times per ensemble.
We are taking this issue into account  in our future software development, by improving and enabling linking to state-of-the-art
QCD software libraries within tmLQCD, such as QUDA, as well as by developing a new user-friendly flexible
python API \textit{lyncs} \cite{Bacchio:lattice}. This will enable HMC simulations on the next generation of high performance systems.

\section{Acknowledgments}

We thank all members of the ETM collaboration for a most conducive cooperation. J.F.~and S.B.~are supported by the H2020 project PRACE 6-IP (GA No. 82376) and the EuroCC (GA No. 951740). P. D.~acknowledges support from the European Unions Horizon 2020 research and innovation programme under the Marie Sk\l{}odowska-Curie grant agreement No.~813942 (EuroPLEx) and from INFN under the research project INFN-QCDLAT. G.K.~acknowledges support from project NextQCD, co-funded by the European Regional Development Fund and the Republic of Cyprus through the Research and Innovation Foundation (RIF) (EXCELLENCE/0918/0129). Partial support is provided by the H2020 European Joint Doctorate program STIMULATE grant No. 765048. We acknowledge the Gauss Centre for Supercomputing e.V. (www. gauss-centre.eu) for project pr74yo by providing computing time on SuperMUC at LRZ (www.lrz.de). The authors acknowledge the Texas Advanced Computing Center (TACC) at The University of Texas at Austin for providing HPC resources that have contributed to the research results reported within this proceeding. This work used resources from NIC on JUWELS and Jureca Booster at the JSC, under projects with ids ECY00, HBN28 and HCH02. We acknowledge PRACE for awarding us access to  HAWK at HLRS, where part of our work is carried out within the project with Id Acid 4886.

\end{document}